\documentclass[final,1p,times]{elsarticle}

\journal{Nuclear Physics A}

\begin{document}

\begin{frontmatter}


\title{Single-pole nature of $\Lambda (1405)$ and structure of $K^-pp$}

\author[label1,label2]{Yoshinori Akaishi}
\ead{akaishi@post.kek.jp}
\author[label1,label3]{Toshimitsu Yamazaki}
\author[]{Mitsuaki Obu}
\author[label2]{Masanobu Wada}

\address[label1]{RIKEN Nishina Center, Wako, Saitama 351-0198, Japan}
\address[label2]{College of Science and Technology, Nihon University, Funabashi, Chiba 274-8501, Japan}
\address[label3]{Department of Physics, University of Tokyo, Tokyo 113-0033, Japan}

\begin{abstract}

We have studied the structure of $K^- pp$ by solving this system in a variational treatment, starting from Ansatz that $\Lambda(1405)$ is a $K^-p$ quasi-bound state, $\Lambda^*$ with mass 1405 MeV/$c^2$. The structure of $K^-pp$ reveals a molecular feature, namely, the $K^-$ in an ``atomic center", $\Lambda^*$, plays a key role in producing strong covalent bonding with the other proton. Deeply bound $\bar K$ nuclear systems are formed by this "super-strong" nuclear force due to migrating real bosons, $\bar K$, {\it a la} Heitler-London-Heisenberg, which overcompensates the stiff nuclear incompressibility. Theoretical background of the $\Lambda (1405)$ Ansatz is discussed in connection with the double-pole picture of $\Lambda (1405)$ based on chiral SU(3) dynamics. Detailed analysis reveals single-pole nature of the observable $\Lambda (1405)$. There are two kinds of $\Sigma \pi$ invariant masses experimentally observable, the usual $T_{22}$ invariant mass and the conversion $
 T_{21}$ invariant mass.  It is of vital importance to determine whether the $\Lambda^*$ mass is 1405 MeV or 1420 MeV. The $T_{21}$ invariant mass from $K^-$ absorption at rest in deuteron can provide decisive information about this $\Lambda^*$ mass problem.

\end{abstract}


\end{frontmatter}



\section{Introduction}	

Traditionally, the $\Lambda(1405)$ resonance has been interpreted as a $K^- p$ quasi-bound state embedded in the $\Sigma \pi$ continuum \cite{Dalitz59}. We have predicted deeply bound kaonic states with large nuclear densities, and studied their structure and formation \cite{Akaishi02,Yamazaki02,Dote04,Yamazaki04,Akaishi05,Kienle06,Yamazaki07} using the $\bar KN$ interaction, which was derived so as to account for the empirically known low-energy $\bar KN$ quantities. This strong binding scheme is shown to originate from the most basic system $K^- pp$ in which the $K^-$ migrates between two protons forming a dense molecular structure with $\Lambda(1405)$ as a basic constituent, and to cause super-strong nuclear force \cite{Yamazaki07b}. Faddeev calculations of $K^-pp$ pole energy \cite{Shevchenko07,Ikeda07,Shevchenko07b} support our original prediction of the possible existence of deeply bound $K^-pp$.

Recently, a totally different theoretical framework with a double-pole structure of $\Lambda (1405)$ has emerged on the basis of chiral SU(3) dynamics, which claims that the $K^- p$ state should exist as a shallow bound state around 1420 MeV \cite{Magas05,Hyodo08}, not the deep one around 1405 MeV. This weak binding scheme necessarily results in predictions of shallow kaonic bound states \cite{Dote09}. Thus, before going to problems of kaonic nuclear systems \cite{Oset06,Yamazaki07c}, it becomes urgently important to ask the question: where is the $K^-p$ bound state? Besides theoretical debates, partly given in this paper, on the above controversial situation, we examine the two scenarios based on experimentally observable $\Sigma \pi$ invariant masses from $K^-$$^4$He and $K^-D$ atoms. In order to check the $\Lambda (1405)$ Ansatz, a $\chi^2$ analysis of $\Sigma \pi$ invariant-mass data from stopped $K^-$ on $^4$He has been performed by calculating theoretical spectra for an
 y value of assumed $\Lambda^*$ mass \cite{Esmaili09}. The $K^-D$ case is more interesting, since $D$ is characterized by not only its long tail but also sizable short-range correlation due to a strong repulsion of $pn$ interaction. It is shown with theoretical foundation that a precise data from stopped $K^-$ on $D$ can distinguish the traditional $\Lambda(1405)$ Ansatz around 1405 MeV from the new claims of less bound $K^-p$ around 1420 MeV. 

\section{Structure of $K^- pp$ and super-strong nuclear force} \label{struc}

Three-body variational wave function of $\bar KNN$ with $(1, 2, 3) = (\bar K, N, N)$ labeling is given in the ATMS method \cite{Akaishi86} as 
\begin{equation}
\Psi = [~\Phi_{12} + \Phi_{13}~]\,\,|T=1/2\rangle ,
\end{equation}
where
\begin{eqnarray} \label{eq:Phi}
\Phi_{12} = [f^{I=0} (r_{12})\, P_{12}^{I=0}  + f^{I=1} (r_{12})\, P_{12}^{I=1}]~f_{NN} (r_{23}) f(r_{31}),\\
\Phi_{13} = [f^{I=0} (r_{31})\, P_{31}^{I=0}  + f^{I=1} (r_{31})\, P_{31}^{I=1}]~f (r_{12}) f_{NN}(r_{23})~, 
\end{eqnarray}
with isospin projection operators, $P_{12}^{I=0}$ and $P_{12}^{I=1}$.
The functions $f^{I=0} (r_{ij})$ and $f^{I=1} (r_{ij})$ are two-body correlation functions of the particle pair $(i,j)$ for the $I=0$ and $I=1$ $\bar KN$ states, respectively, and $f_{NN} (r_{23})$ is that for the $NN$ pair, and $f(r_{i,j})$ is for highly off-shell $\bar KN$ cases.  The $T=1/2$ state consists of two isospin eigenstates as
\begin{equation}
|T=1/2\rangle =  \sqrt{\frac{3}{4}} \, \biggl[(\bar{K}_1 N_2)^{0,0} \, p_3 \biggr] + \sqrt{\frac{1}{4}} \, \biggl[ - \sqrt{\frac{1}{3}} (\bar{K}_1 N_2)^{1,0} \, p_3 
+ \sqrt{\frac{2}{3}} (\bar{K}_1 N_2)^{1,1} \, n_3 \biggr] , 
\end{equation}
where $(\bar{K}_1 N_2)^{I, I_z}$ is for the isospin $(I, I_z)$ state. Among these the first term corresponds to $\Lambda^*$-$p$ structure.
We use the single-channel effective $\bar KN$ potentials with imaginary parts in energy-independent form, which is an appropriate way to obtain the decaying state of Kapur-Peierls \cite{Kapur38} as discussed in Ref. \cite{Akaishi08}. The complex $\bar KN$ potentials are:
\begin{eqnarray}
v_{\bar{K}N}^{I=0} (r) = (-595~ - { i} \, 83)\, {\rm exp} [-(r/0.66)^2], \label{vKN0} \\
v_{\bar{K}N}^{I=1} (r) = (-175 - { i} \, 105)\, {\rm exp} [-(r/0.66)^2],\label{vKN1}
\end{eqnarray}
in units of MeV and fm, and Tamagaki's potential is employed as realistic $NN$ interaction.

\begin{figure}
\centering
\includegraphics[width=10cm]{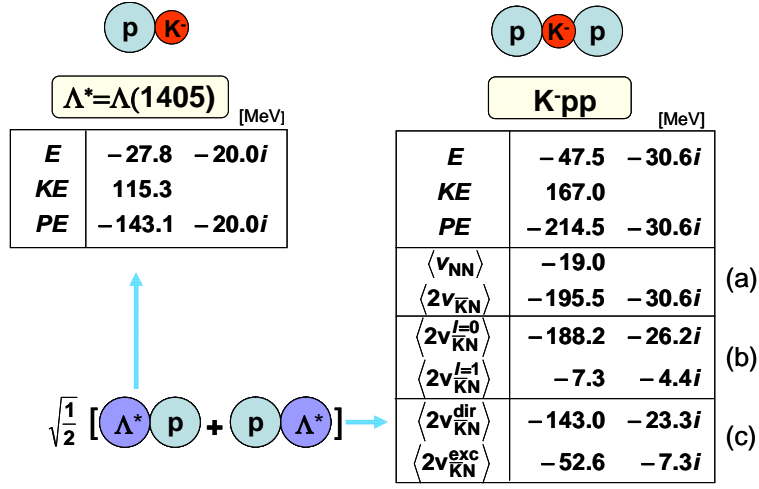}
\vspace{0.cm}
\caption{\label{energy} 
Energy breakdown of $K^-pp$ together with that of $\Lambda ^* = \Lambda (1405)$. The binding energy is a result of large cancellation between kinetic and potential energies. Potential energy contributions are compared (a) between $NN$ and $\bar KN$ interactions, (b) between $I=0~ \bar KN$ and $I=1~ \bar KN$ interactions and (c) between diagonal and exchange matrix-elements of the $\bar KN$ interaction. }
\end{figure}

The details of energies obtained for $K^-pp$ are shown in Fig.~\ref{energy} together with those of $\Lambda ^* = \Lambda (1405)$. In spite of unbound $pp$ the $K^-$ can combine the two protons into a quasi-bound system, $K^-pp$.  
%
It is interesting to analyze what the individual energy term means in the light of the Heitler-London picture \cite{Heitler27,Heisenberg32}. The "atomic" system, $K^-p$, has total energy of $-27.8 - i 20.0$ MeV, kinetic energy of 115.3 MeV and potential energy of $-143.1 - i 20.0$ MeV. Those energies of the "molecular" system, $K^-pp$, are $-47.5 - i 30.6$ MeV, 167.0 MeV and $-214.5 - i 30.6$ MeV, respectively. The potential energy is classified in three ways in Fig.~\ref{energy}. The column (a) shows that about 90\% comes from $\bar KN$ interaction, and the column (b) indicates that the $I=0~\bar KN$ interaction markedly dominates, suggesting a $\Lambda ^*$ cluster structure in the $K^-pp$ system. The column (c) divides the $\bar KN$ contribution into diagonal and exchange integrals of the Heitler-London picture. The diagonal part is surprisingly close to the potential energy of free $\Lambda ^*$, confirming the $\Lambda ^*$-$p$ structure of $K^-pp$. The exchange part is ess
 entially important as discussed just below.

It is emphasized that the strong $I=0~\bar KN$ attraction produces a large exchange integral,
\begin{equation}
\sum _{(i,j) = (2,3),(3,2)} \langle \Phi_{1i} |v_{\bar KN} (12) + v_{\bar KN} (13) |\Phi_{1j}\rangle  = -52.6 - i 7.3~{\rm MeV},
\end{equation}
which is the source for the deeper binding of $K^-pp$ as compared with the $\Lambda^*+p$ threshold.
Thus, the $I=0~\bar KN$ exchange attraction produces a very strong molecular type bonding between the two protons. The molecular $K^-pp$ state resembles a state, $[(\Lambda^*$-$p) + (p$-$\Lambda^*)]~/\sqrt 2$, tightly bound by exchange of a real $\bar K$. This adiabatic $pp$ potential due to the migration of $\bar K$ is called {\it super-strong nuclear force} \cite{Yamazaki07b}, which is about 4-times stronger than the ordinary $NN$ force.

\section{Single-pole nature of $\Lambda (1405)$} \label{Singl}

\subsection{$\bar KN$-$\Sigma \pi$ coupled-channel system}

We treat the $K^-p$ quasi-bound state as a Feshbach resonance \cite{Feshbach58} embedded in the $\Sigma \pi$ continuum by using Akaishi-Myint-Yamazaki's (AMY) phenomenological model \cite{Akaishi08} and also Hyodo-Weise's (H-W) two-channel model of chiral SU(3) dynamics \cite{Hyodo08}. 
In the AMY model, we employ a set of separable potentials with a Yukawa-type form factor for the coupled system of $\bar KN$ and $\Sigma \pi$ channels,
\begin{eqnarray}
\langle \vec k'_i \mid v_{ij} \mid \vec k_j \rangle = g(\vec k'_i) ~U_{ij}~ g(\vec k_j), ~~~g(\vec k) = \frac {\Lambda^2} {\Lambda^2 + \vec k^2}, 
\label{KNint}
\end{eqnarray}
where we impose a constraint of $U_{\Sigma\pi,\Sigma\pi}/U_{\bar KN,\bar KN}= 4/3$ and take $\Lambda$ to be 3.90 fm$^{-1}$. In this model the loop integral is calculated to be
\begin{equation}
G_j = -\pi ^2 \frac{2\mu_j}{\hbar^2} \Lambda \left( \frac{\Lambda}{\Lambda -i k_j} \right)^2, 
\end{equation}
where $\mu_j$ is the reduced mass and $k_j$ is a relative momentum in the channel $j$. The H-W case can be treated in the same framework by using the Weinberg-Tomozawa term as $U_{ij}$ of zero-range ($\Lambda=\infty$) and the regularized $G_j$ of Eq. (3) in Ref. \cite{Hyodo08}.

The transition-matrix of the two coupled channels, $\bar KN (1)$ and $\Sigma \pi (2)$, obeys the following equation; 
\begin{equation}
\left( \begin{array}{cc}
T_{11} & T_{12} \\
T_{21} & T_{22} 
\end{array} \right)  
= 
\left( \begin{array}{cc}
U_{11} & U_{12} \\
U_{21} & U_{22} 
\end{array} \right) 
+ 
\left( \begin{array}{cc}
U_{11} & U_{12} \\
U_{21} & U_{22} 
\end{array} \right)
\left( \begin{array}{cc}
G_{1} & 0 \\
0 & G_{2} 
\end{array} \right)
\left( \begin{array}{cc}
T_{11} & T_{12} \\
T_{21} & T_{22} 
\end{array} \right) . 
\label{cceq}
\end{equation}
The solutions of each matrix element are {\it exactly} given by
\begin{equation}
T_{ii} = \frac {1}{1-U^{\rm{opt}}_{ii} G_i} U^{\rm{opt}}_{ii},~~~
T_{ji} = \frac {1}{1-U^{\rm{opt}}_{jj} G_j} U^{\rm{opt}}_{ji}
\label{sceq}
\end{equation}
with generalized optical potentials, 
\begin{equation}
U^{\rm{opt}}_{ii} = U_{ii} + U_{ij} \frac {G_j}{1-U_{jj} G_j} U_{ji},~~~
U^{\rm{opt}}_{ji} = U_{ji} \frac {1}{1-U_{ii} G_i},
\end{equation}
where $(i,j)=(1,2),(2,1)$.
It should be noticed that the two-channel coupled equation is divided into four single-channel effective equations without any approximation by the use of the optical potentials. 

Among the matrix elements, $T_{11},~T_{12},~T_{21}$ and $T_{22}$, the experimentally observable quantities below the $\bar K+N$ threshold are $- \frac {1} {\pi} ~{\rm Im}~ T_{11}, ~| T_{21} |^2 k_2$ and $| T_{22} |^2 k_2$, where $k_2$ is a $\Sigma \pi$ relative momentum. The first one is a $\bar KN$ missing-mass spectrum, and is proportional to the imaginary part of the $\bar KN$ scattering amplitude, the peak position of which is just of our concern. The second one is a $\Sigma \pi$ invariant-mass spectrum from the conversion process, $\bar KN \rightarrow \Sigma \pi$ (we call this "$T_{21}$ invariant mass"). The third one is a $\Sigma \pi$ invariant-mass spectrum from the scattering process, $\Sigma \pi \rightarrow \Sigma \pi$ (we call this "$T_{22}$ invariant mass"). The $T_{21}$ invariant-mass spectrum coincides with the $\bar KN$ missing-mass spectrum in the mass region below the $\bar K$+$N$ threshold: see Eq. (\ref{opTheor}). 

\subsection{$T_{22}~ \Sigma \pi$ invariant-mass spectrum}

The "double-pole structure" of $\Lambda (1405)$ in chiral SU(3) dynamics has been revealed for the first time by Jido {\it et al.} \cite{Jido03}: one pole (we refer to it as the 1st pole) appears at $-1426 - i 16$ MeV and the other (2nd pole) at $-1390 - i 66$ MeV. The $T_{22}~ \Sigma \pi$ invariant mass has a peak at around 1405 MeV, which may be explained by a superposition of the two resonance amplitudes having the above poles. We critically examine this "double-pole explanation" of $\Lambda (1405)$. 

\begin{figure}
\centering
\includegraphics[width=12cm]{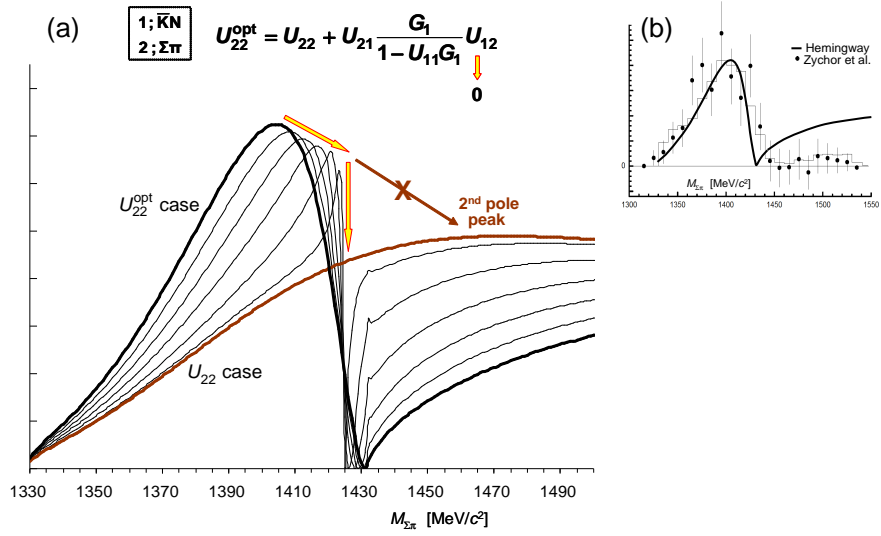}
\vspace{0.cm}
\caption{\label{T22} 
$T_{22} ~\Sigma \pi$ invariant-mass spectrum calculated by H-W's two-channel treatment of chiral SU(3) dynamics. (a) Change of spectrum shape when the coupling to the $\bar KN$ channel is reduced: the peak goes to disappear at the 1st pole position and never approaches to the 2nd pole peak. (b) The H-W spectrum is compared with experimental data of Hemingway \cite{Hemingway85} and of Zychor {\it et al.} \cite{Zychor08}. A large discrepancy is seen in the region above the $\bar K$+$N$ threshold. }
\end{figure}

Figure \ref{T22} shows a $T_{22}~ \Sigma \pi$ invariant-mass spectrum calculated by Hyodo-Weise (H-W)'s two-channel treatment of chiral SU(3) dynamics. H-W give the 1st pole at $1432 - i 17$ MeV and the 2nd pole at $1398 - i 73$ MeV. In order to disclose the respective roles of the 1st and the 2nd poles we divide the $\Sigma \pi$ interaction as, 
\begin{equation}
U^{\rm{opt}}_{22} = U_{22} + f \times U_{21} \frac {G_1}{1-U_{11} G_1} U_{12} 
\label{U22}
\end{equation}
with a reduction factor $f$ on the coupling term, and investigate the effects of each term of the right-hand side on the spectrum. 
The first term, $U_{22}$, is the main origin of the 2nd pole, giving a pole at $1388 - i 96$ MeV in only $\Sigma \pi$ channel treatment. This resonance-pole appearance is due to a strong energy-dependence of the $\Sigma \pi$ interaction, especially due to its {\it positive} imaginary part induced through a self-consistent complex $\sqrt s -M_{\Sigma}$ in the Weinberg-Tomozawa term at the pole position. However, the curve on real $\sqrt s$ axis, denoted as "$U_{22}$ case" in Fig. \ref{T22}, has no peak structure at 1388 MeV but shows a broad bump around 1470 MeV far above the $\bar K$+$N$ threshold. AMY \cite{Akaishi08} discussed that experimental observation corresponds not to a pole state but to a decaying state, since detectable decay particles appear as on-shell objects in their asymptotic region. The decaying state of $U_{22}(\rm{real} \sqrt s)$ is very different from the pole state of $U_{22}(\rm{complex} \sqrt s)$ for the strongly energy-dependent $\Sigma \pi$ interacti
 on. It should be noticed that the "two-pole superposition" is not a reasonable explanation of $\Lambda (1405)$, since the 1st and 2nd poles give peaks around 1420 MeV and 1470 MeV respectively, both of which are higher than 1405 MeV.

The second term of Eq. (\ref{U22}) gives the contribution from the 1st pole. When we reduce the strength of this term as $f = 0.8, 0.6, 0.4, 0.2 ~\rm{and}~ 0.1$, the peak at 1405 MeV (1405 $T_{22}$ peak) of "$U^{\rm{opt}}_{22}$ case", that is exactly of H-W's two-channel system, converges to the 1st-pole position and disappears, and never approaches to the peak of "$U_{22}$ case". This fact clearly shows that the 1405 $T_{22}$ peak is of the 1st-pole origin. This peak structure is a result of interference between sharp resonance amplitude from the 1st pole and continuum amplitude slowly increasing toward the maximum around 1470 MeV, which is nothing but the contribution of the 2nd pole on the real $\sqrt s$ axis. Thus, the 2nd pole is irrelevant to any peak structure in the mass region between the $\Sigma$+$\pi$ and the $\bar K$+$N$ thresholds. Magas {\it et al.} \cite{Magas05} claimed an experimental evidence of "double-pole structure" of $\Lambda (1405)$, but it is logicall
 y impossible to stand since the 1405 $T_{22}$ peak is a remnant of the 1st pole as shown here. 
From the above consideration we can conclude that {\it the observable $\Lambda (1405)$ is of single-pole nature}.  

\begin{figure}
\centering
\includegraphics[width=12cm]{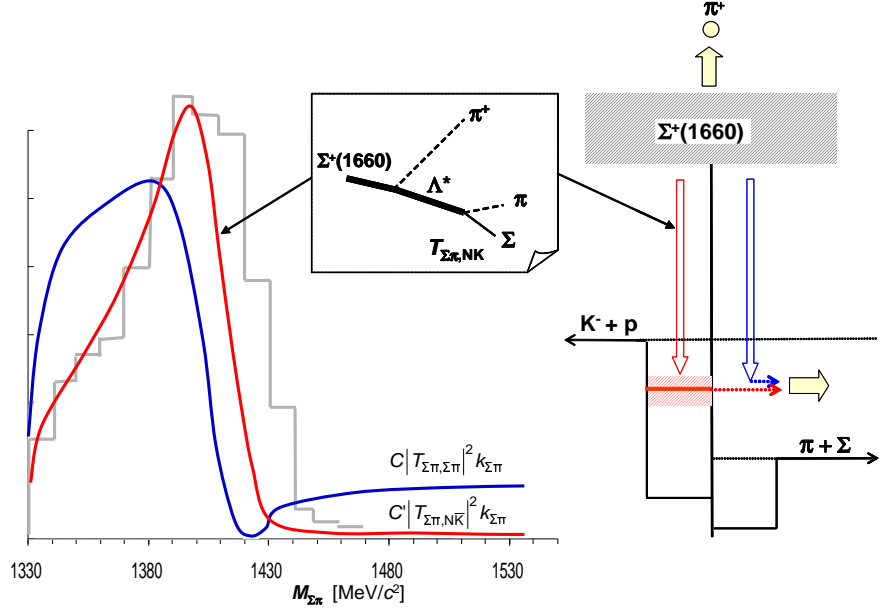}
\vspace{0.cm}
\caption{\label{Spi} 
$\Sigma^+ \pi^-$ invariant mass spectrum. A histogram is the experimental data of Hemingway \cite{Hemingway85}. Theoretical curves are calculated with AMY interaction by assuming the $\Lambda (1405)$ mass to be 1405 MeV. The $T_{22}$ fitting is miserable, but the $T_{21}$ fit gives a good result, where 1 and 2 stand for the $\bar KN$ and the $\Sigma \pi$ channels, respectively.}
\end{figure}

Now, our problem is to discriminate whether the position of the single-pole $\Lambda (1405) = \Lambda^*$ is 1420 MeV or 1405 MeV. 
One of experimental data available for $\Lambda^*$ mass determination was provided by Hemingway \cite{Hemingway85}, which is shown with a histogram in Fig.~\ref{Spi} together with his sketch of the formation-to-decay process of $\Lambda^*$. This $\Sigma^+ \pi^-$ data has been believed to be fitted with $T_{\Sigma \pi,\Sigma \pi}~ (T_{22})$ matrix element. In Fig. \ref{Spi} theoretical curves are calculated with AMY interaction by assuming the $\Lambda^*$ pole at $1405 - i 20$ MeV. The $T_{22}$ fitting is poor, but the $T_{21}$ fit, which has not been considered so far, gives a rather better result. However, if we assume the $\Lambda^*$ pole at $1420 - i 17$, the usual $T_{22}$ fit gives better result. 

In the case of H-W's chiral SU(3) dynamics, as shown in Fig. \ref{T22}(b), the $T_{22}$ invariant-mass spectrum well reproduces the $\Sigma^+ \pi^-$ data of Hemingway and also the $\Sigma^0 \pi^0$ data of Zychor {\it et al.} \cite{Zychor08} below the $\bar K$+$N$ threshold. The chiral theory, however, largely overshoots the data above the $\bar K$+$N$ threshold due to 1.7-2.6 times stronger Weinberg-Tomozawa term in this mass region than that at the $\Sigma$+$\pi$ threshold.  On the other hand, Geng-Oset \cite{Geng07} applied successfully a $T_{21}$ fit to Zychor {\it et al.}'s data. Thus, we have to solve the "$T_{22}$ or $T_{21}$ (or their mixing) fit" problem before to draw any conclusion about the $\Lambda (1405)$ mass from these data.

\subsection{$T_{21}~ \Sigma \pi$ invariant-mass spectrum}

Few-body kaonic atoms have an advantage that the $T_{21}$ fitting is specified. We investigate $T_{21} ~\Sigma \pi$ invariant-mass spectra from $K^-$ absorption in $^4$He and in $D$. One may think that these spectra come from quasi-free decay processes, but that is not true: all the spectra projected with the respective spectator momenta are essentially the conversion $\Sigma \pi$ spectra via the $\Lambda (1405)$ quasi-bound state. Esmaili {\it et al.} performed a $\chi^2$-fit analysis of the $K^{-4}$He bubble chamber data given by Riley {\it et al.} \cite{Riley75}. The $\chi^2$ contour map calculated for separable interactions of Eq. (\ref{KNint}) with variable mass, $M$, and width, $\Gamma$, has determined the best-fit value \cite{Esmaili09} to be
\begin{equation}
M = 1405.5^{+1.4}_{-1.0}~{\rm MeV}/c^2~ {\rm and}~ \Gamma = 25.6 ^{+4}_{-3}~{\rm MeV}. 
\label{best}
\end{equation}
This result strongly supports our $\Lambda (1405)$ Ansatz of 1405 MeV: the 1420 MeV mass of H-W is located outside of 99.9\% confidence level in the $\chi^2$ map. 

\begin{figure}
\centering
\includegraphics[width=12cm]{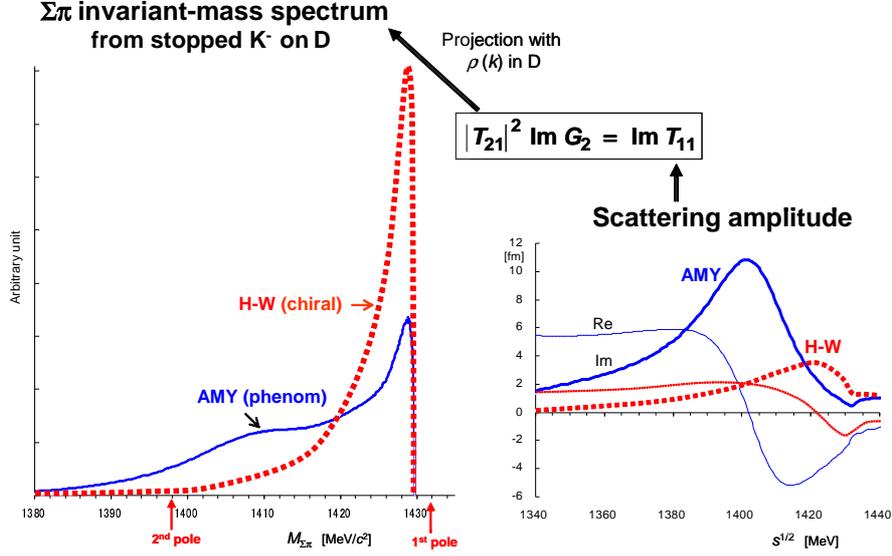}
\vspace{0.cm}
\caption{\label{KDatom} 
$T_{21} ~\Sigma \pi$ invariant-mass spectra from $K^-D$ atom and imaginary parts of $\bar KN$ scattering amplitude, both of which are calculated for H-W and AMY cases. The measurement of $\Sigma \pi$ invariant mass is virtually equivalent to the measurement of the imaginary part of the scattering amplitude.}
\end{figure}

The most important advantage of kaonic-atom absorption process comes from the following optical relation.
From Eq. (\ref{sceq}) we can prove that the relation,
\begin{equation}
{\rm Im}~ T_{11} = |T_{21}|^2 ~{\rm Im} ~G_2, 
\label{opTheor}
\end{equation}
holds below the $\bar K$+$N$ threshold, where Im $G_2$ is proportional to phase volume of the decay channel. This equality means that the observation of the $T_{21}$ invariant-mass spectrum is nothing but the observation of the imaginary part of the scattering amplitude given in Fig. 15 of Hyodo-Weise \cite{Hyodo08}, from which the less bound $K^-p$ around 1420 MeV has been claimed against the traditional $\Lambda(1405)$ Ansatz around 1405 MeV.  
Figure \ref{KDatom} shows $T_{21} ~\Sigma \pi$ invariant-mass spectra from $K^-D$ atom and imaginary parts of $\bar KN$ scattering amplitude, both of which are calculated for H-W and AMY cases by Esmaili {\it et al.} \cite{Esmaili09a}. The experimental feasibility has been investigated by Suzuki {\it et al.} \cite{Suzuki09}.

\section{Conclusions} \label{Conc}

The $\Lambda (1405)$ quasi-bound state forms the basic structure of the kaonic nuclear cluster $K^-pp$, which can be interpreted as a "kaonic hydrogen molecule". The migrating $K^-$ produces "strong covalency" between the two protons through the strongly attractive $I=0~\bar KN$ interaction. This is essentially the mechanism of Heitler and London for hydrogen molecule, though the nature of the interaction is totally different and the migrating particle is a much heavier boson. This is a revival of Heisenberg picture. Despite the drastic dynamical change of the system caused by the strong $\bar K N$ interaction, the identity of the "constituent atom", $\Lambda^*$, is nearly preserved. 

The observable $\Lambda (1405)$ is of single-pole nature, since the 2nd pole in chiral SU(3) dynamics, which appears due to strong energy-dependence of the interaction, is irrelevant to any experimental peak structure. It is important to distinguish the pole position of $\Lambda (1405)$, 1405 MeV or 1420 MeV, by considering the "$T_{21}/T_{22}$" fitting ambiguity. The $\Sigma \pi$ invariant-mass spectrum from few-body kaonic atoms has an advantage that the entrance channel is uniquely determined. A conversion ($T_{21}$) $\Sigma \pi$ invariant-mass spectrum from stopped $K^-$ absorption in $^4$He favors our $\Lambda (1405)$ Ansatz, giving the best-fit value of Eq. (\ref{best}). Stopped $K^-$ on $D$ would provide a decisive datum concerning the most urgent problem of the $\Lambda (1405)$ mass \cite{Esmaili09a,Suzuki09}. Such experiments at J-PARC and others are highly awaited.
\\

We thank Mr. Jafar Esmaili and Prof. Khin Swe Myint for collaborations and discussions.

\end{document}